%% file: report.tex
\documentclass[10pt,twoside]{report}
\usepackage{url}
\usepackage{graphicx}
\usepackage{caption}
\usepackage{subcaption}
\usepackage{listings}
\usepackage{xcolor}
\usepackage{framed}
\usepackage{pgfplotstable}
\usepackage{pgfplots}
\usepackage{color}
\usepackage{amsmath}

\newcommand{\cmnt}[1]{}

\newcommand{\Alpha}[1]{\ensuremath{\alpha(#1)}}
\newcommand{\Beta}[1]{\ensuremath{\beta(#1)}}
\newcommand{\myGamma}[1]{\ensuremath{\gamma(#1)}}
\newcommand{\myin}[1]{\texttt{In(#1)}}
\newcommand{\myout}[1]{\texttt{Out(#1)}}
\newcommand{\antin}[1]{\texttt{Antin(#1)}}
\newcommand{\antout}[1]{\texttt{Antout(#1)}}
\newcommand{\antloc}[1]{\texttt{Antloc(#1)}}
\newcommand{\transp}[1]{\texttt{Tranp(#1)}}
\newcommand{\xcomp}[1]{\texttt{Xcomp(#1)}}
\newcommand{\availin}[1]{\texttt{Availin(#1)}}
\newcommand{\availout}[1]{\texttt{Availout(#1)}}
\newcommand{\earlin}[1]{\texttt{Earliestin(#1)}}
\newcommand{\earlout}[1]{\texttt{Earliestout(#1)}}
\newcommand{\delayin}[1]{\texttt{Delayin(#1)}}
\newcommand{\delayout}[1]{\texttt{Delayout(#1)}}
\newcommand{\latestin}[1]{\texttt{Latestin(#1)}}
\newcommand{\latestout}[1]{\texttt{Latestout(#1)}}
\newcommand{\isoin}[1]{\texttt{Isolatedin(#1)}}
\newcommand{\isoout}[1]{\texttt{Isolatedout(#1)}}
\newcommand{\insertin}[1]{\texttt{Insertin(#1)}}
\newcommand{\insertout}[1]{\texttt{Insertout(#1)}}
\newcommand{\replacein}[1]{\texttt{Replacein(#1)}}
\newcommand{\replaceout}[1]{\texttt{Replaceout(#1)}}
\title{\textbf{Partial Redundancy Elimination using Lazy Code Motion}}

\author{Sandeep Dasgupta\thanks{Electronic address:
\texttt{sdasgup3@illinois.edu}} \qquad Tanmay Gangwani\thanks{Electronic
address: \texttt{gangwan2@illinois.edu}}}

\begin{document}
\begin{titlepage}
\thispagestyle{empty}
\maketitle
\pagebreak
\end{titlepage}

\begin{flushleft}
\textbf{\Large{1. Problem Statement \& Motivation}}
\end{flushleft}
Partial Redundancy Elimination (PRE) is a compiler optimization that eliminates
expressions that are redundant on some but not necessarily all paths through a
program. In this project, we implemented a PRE optimization pass in LLVM 
and measured results on a variety of applications. We chose PRE because
it's a powerful technique that subsumes Common Subexpression Elimination 
(CSE) and Loop Invariant Code Motion (LICM), and hence has a potential to 
greatly improve performance.

In the example below, the computation of the expression (a + b) is partially
redundant because it is redundant on the path $1 \rightarrow 2 \rightarrow 5$,
          but not on the path $1 \rightarrow 4 \rightarrow 5$. PRE works by first introducing
          operations that make the partially redundant expressions fully
          redundant and then deleting the redundant computations. The
          computation of (a + b) is added to 4 and then deleted from 5.

\lstset{ %
  basicstyle=\footnotesize, 
  breakatwhitespace=false,   
  breaklines=true,            
}
\begin{center}
\begin{lstlisting}
    (1) if (OPAQUE)
    (2)   x = a + b;
    (3) else
    (4)   x = 0;
    (5) y=a+b;
\end{lstlisting}
\end{center}

\begin{flushleft}
\textbf{\Large{2. Related Work}}
\end{flushleft}

\begin{description}
\item [Partial Redundancy Elimination]    \hfill \\
Morel et al. \cite{Morel} first proposed a bit-vector algorithm for the
suppression of partial redundancies. The bi-directionality of the algorithm,
            however, proved to be computationally challenging. Knoop et al.
            \cite{Knoop} solved this problem with their Lazy Code Motion (LCM)
  algorithm. It is composed of uni-directional data flow equations and provides
  the earliest and latest placement points for operations that should be
  hoisted. Drechsler et al. \cite{Drechsler} present a variant of LCM which they claim to
  be more useful in practice. Briggs et al. \cite{Briggs2} allude to two
  pre-passes to make PRE more effective - Global Reassociation and Value
  Numbering.

\item [Value Numbering] \hfill \\
Briggs et al. \cite{Briggs} compare and contrast two techniques for
value numbering - hash based\cite{CS} and partition based\cite{AWZ}. In subsequent work they
provide SCC-based Value Numbering \cite{Cooper95scc-basedvalue}  which combines
the best of the previously mentioned approaches. Cooper et al.
\cite{CVDC} show how to incorporate value
information in the data flow equations of LCM to eliminate more redundancies.

\item [PRE in LLVM] \hfill \\
Since LLVM is a Static Single Assignment (SSA) based representation, algorithms
based on identifying expressions which are lexically identical or have the same
static value number may fail to capture some redundancies.
Keneddy Chow et al.
\cite{Kennedy99partialredundancy} provide a new framework for PRE on a program
in SSA form. The present GVN-PRE pass in LLVM appears to be inspired by the
work of Thomas et al. \cite{Vandrunen04value-basedpartial} which also focuses
on SSA.
\end{description}

\begin{flushleft}
\textbf{\Large{3. Algorithm Overview}}
\end{flushleft}

Our algorithm for PRE is a slightly modified version of the iterative bit-vector data flow
algorithm by Knoop et al. \cite{Knoop}. It uses four data flow equations to identify for each expression
in the program, the optimal evaluation point. The first flow equation calculates down-safe (anticipatible) points 
for an expression. An expression is said to be down-safe at a point \emph{p} if computing the 
expression at \emph{p} would be useful along any path from \emph{p}. The second flow equation 
calculates up-safe (available) points. An expression is up-safe at a point \emph{p} is it has been computed
on every path from the entry node to \emph{p} and not killed after the last computation on each 
path. Using these, the algorithm calculates the \emph{Earliest} property. An expression is said to
be \emph{Earliest} at a point \emph{p} if there doesn't exist an earlier point where the computation
of the expression is both down-safe and produces the correct values. Such points are known
as \emph{computationally optimal} placement points. \\

Evaluating the expression at computationally optimal points could negatively impact
performance due to increased register pressure. Therefore, the latter half of the LCM algorithm pushes 
the computation of the expression close to the use of the expression.
More specifically, the third flow equation calculates the \emph{Latest} property. An expression is 
said to be \emph{Latest} at a point \emph{p} if it is computationally optimal at \emph{p}, and on every
path from \emph{p}, any later optimal point on the path would be after some use of the expression.
Through the fourth and final flow equation, the algorithm determines if it is necessary to allocate a 
temporary at a point \emph{p} for the expression. The property is known as \emph{Isolated}. An expression
is \emph{Isolated} at a point if it is optimal, and the value of the expression is only used immediately 
after the point. Therefore, allocation of temporaries at \emph{Isolated} points is avoided.\\

In summary, the four flow equations provide computationally optimal 
placement points which require the shortest lifetimes for the temporary variables introduced. In appendix B, we 
outline all equations.

\begin{flushleft}
\textbf{\Large{4. Implementation Details}}
\end{flushleft}

\begin{flushleft}
\textbf{\large{Value Numbering}}
\end{flushleft}

Prior research \cite{Briggs2} has shown that value numbering can increase opportunities for PRE.
LLVM presently has a GVN-PRE pass which exploits this. However, value numbering in 
GVN-PRE is tightly coupled with the code for removing redundancies, and hence
we were not able to use the same for our code. We wrote our own value numbering
pass which fed expression value numbers to the PRE stage. It should be noted,
however, that we did not implement value numbering from scratch and used an
old (now defunct) LLVM pass as a starting point. Most importantly, we
augmented the basic value numbering in the following ways - 
     
\begin{itemize}     
  \item Added the notion of leader expression (described below), with associated
  data structures and functions. 
  \item Functionality to support value-number-based bitvectors rather than
  expression-name-based bitvectors. 
  \item (Optimization 1) If the expression operator is one of these - AND, OR, CMP::EQ
  or CMP::NE, and the operands have the same value number, we replace all uses
  accordingly and then delete the expression.
  \item (Optimization 2) If all operands of an expression are constants, then we 
  evaluate and propagate constants. 
  \item (Optimization 3) If one operand of an expression is a constant (0 or 1), then 
  we simplify the expression. e.g. {a+0 = a} , {b*1 = b}.
  \item (Optimization 4) If the incoming expressions to a \texttt{Phi} node have the same value 
  number, then the \texttt{Phi} node gets that same value number
\end{itemize}  

  Reassociation has also been shown to make the code more amenable for PRE. It refers to 
using associativity, commutativity and distributivity to divide expressions into parts that are
constant, loop invariant and variable. We used an already existing LLVM pass (-reassociate)
  for Global Reassociation. As per our testing, optimizations 2 and 3 (above) are also done by 
 this pass, and hence, we disabled our version for the more robust LLVM version. Optimizations 1 and 4, 
however, are still our contribution.

\begin{flushleft}
\textbf{\large{Notion Of Leader Expression}}
\end{flushleft}
The value numbering algorithm computes the RPO solution as outlined in
\cite{Cooper95scc-basedvalue}. It goes over the basic blocks in reverse post
order and adds new expressions to a hash table based on the already computed
value numbers of the operands. We call an expression a `leader' if at the time
of computing its value number, the value number doesn't already exist in the
hash table. In other words, out of a potentially large set of expressions that
map to a particular value number, the leader expression was the first to be
encountered while traversing the function in reverse post order. Leader
expressions are vital to our algorithm as they are used to calculate the block
local properties of the data flow equations (Appendix A).

\begin{figure}[htbp]
  \begin{center}
    \scalebox{.6}{\begin{tabular}{@{}p{12cm}@{ } @{ }p{12cm}@{}}
     \includegraphics[scale=0.8]{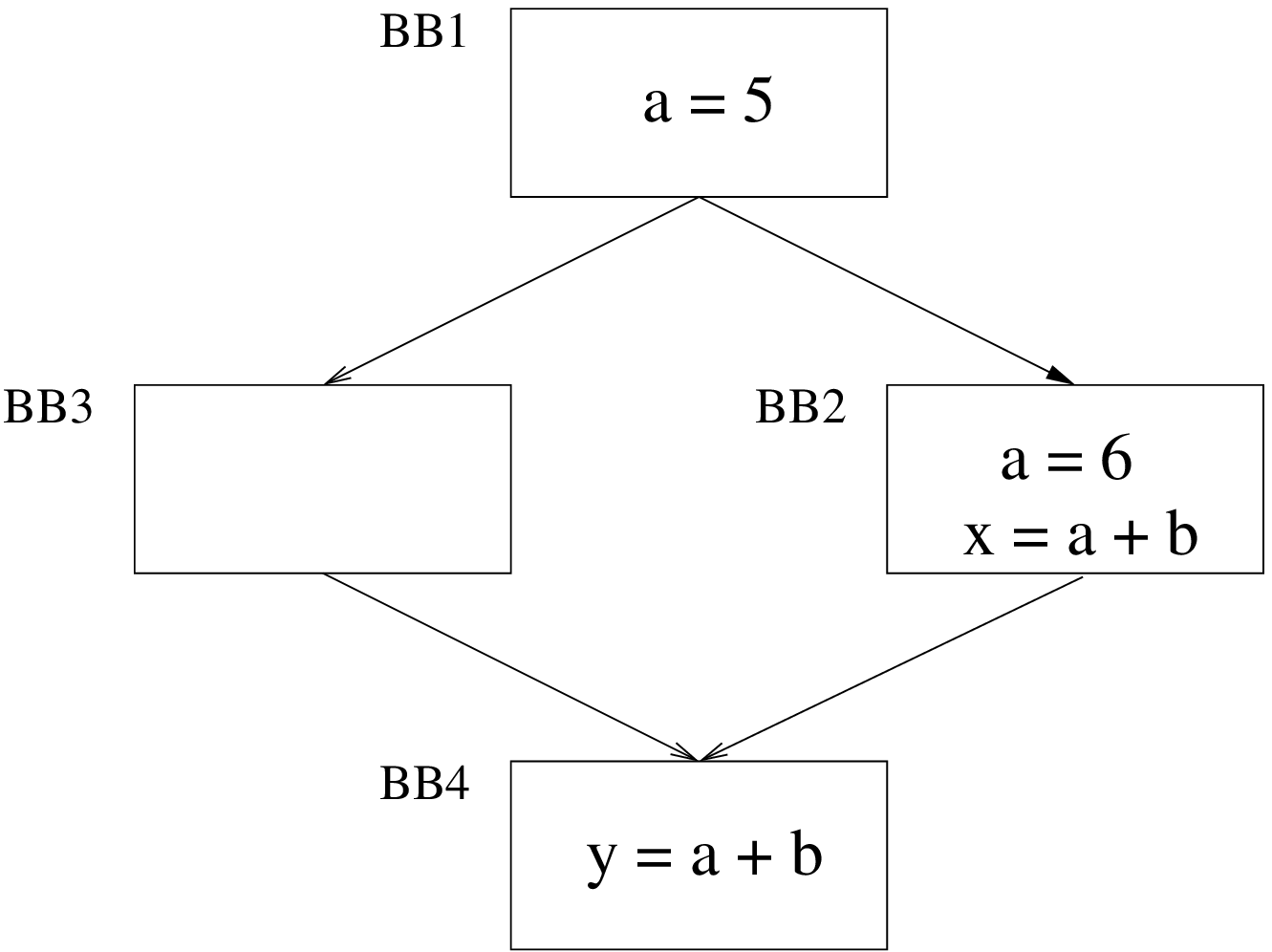} & \includegraphics[scale=0.8]{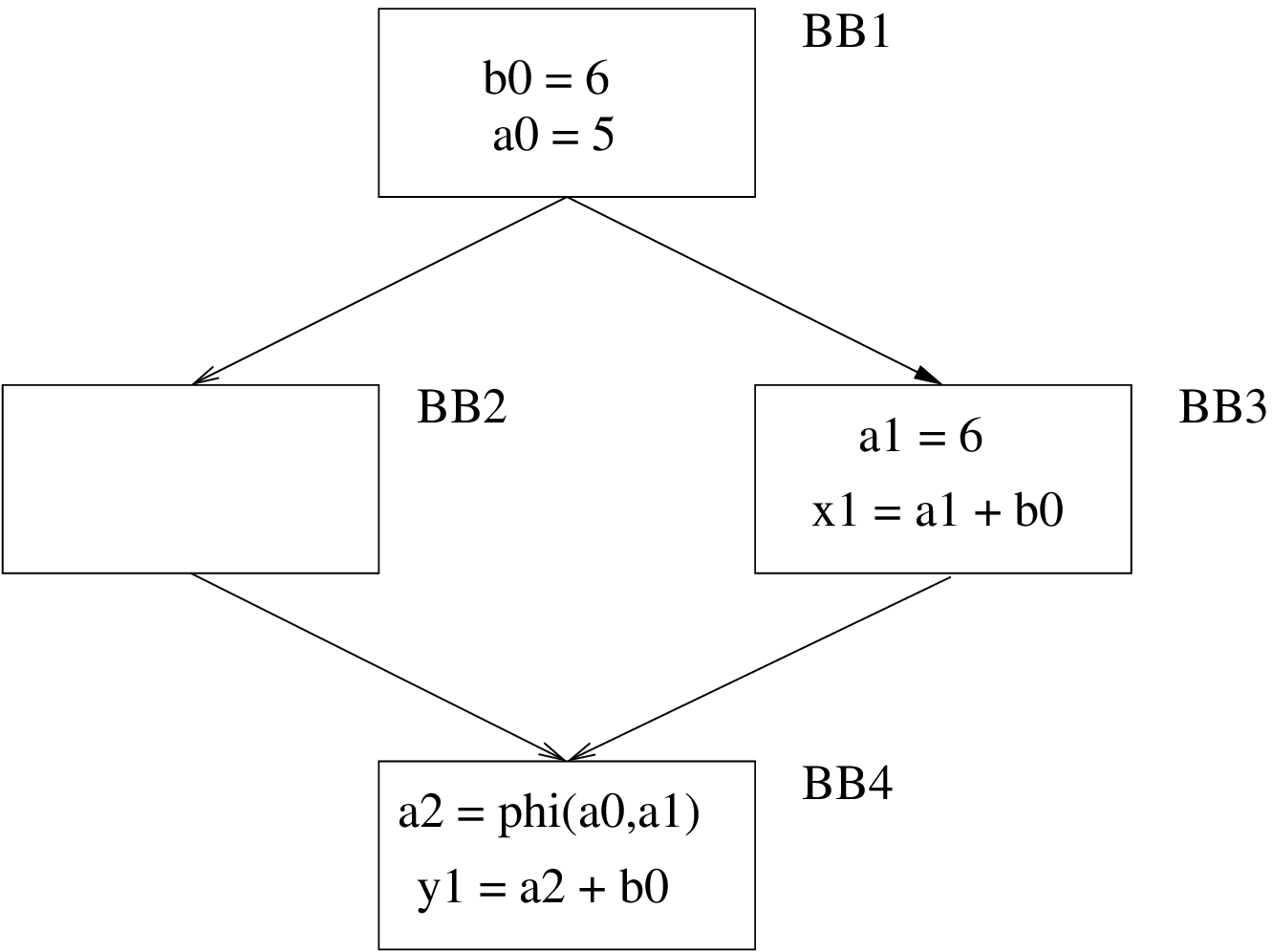} \\
      &\\
  \text{Code not in SSA Form; Two lexically equivalent expressions} &
  \text{Code in SSA Form; Two lexically different expressions} \\
  \text{in Basic block 3 and 4 with different value numbers.} &
  \text{in Basic block 3 and 4 with different value numbers.}
    \end{tabular}}
  \end{center}
  \caption{\label{fig:0} }
\end{figure}

\begin{flushleft}
\textbf{\large{Types Of Redundancies}}
\end{flushleft}
Given two expressions X and Y in the source code, following are the possibilities - \\
\indent1. X and Y are lexically equivalent, and have the same value numbers \\
\indent2. X and Y are lexically equivalent, but have different value numbers \\
\indent3. X and Y are lexically different, but have the same value numbers \\
\indent4. X and Y are lexically different, and have different value numbers \\
In the source code, there could be opportunities for redundancy elimination in
cases 1, 2 and 3 above. If the source code is converted to an intermediate
representation in SSA form then case 2 becomes an impossibility (by guarantees of SSA). Therefore,
               our algorithm presently handles the cases when X and Y are
               lexically same/different, but both have the same value number (cases 1 and 3).
               Driven by this observation, we implement value number based code
               motion, the details of which are presented below. It should be
               noted that even though case 2 above is not possible in SSA,
               the source code redundancies  of this type transform into that
               of type case 4. Figure \ref{fig:0} presents an illustration of the same. \textbf{This is not handled
               in our implementation.} 

\newpage

\begin{figure}[htbp]
\centering
  \includegraphics[scale=0.40]{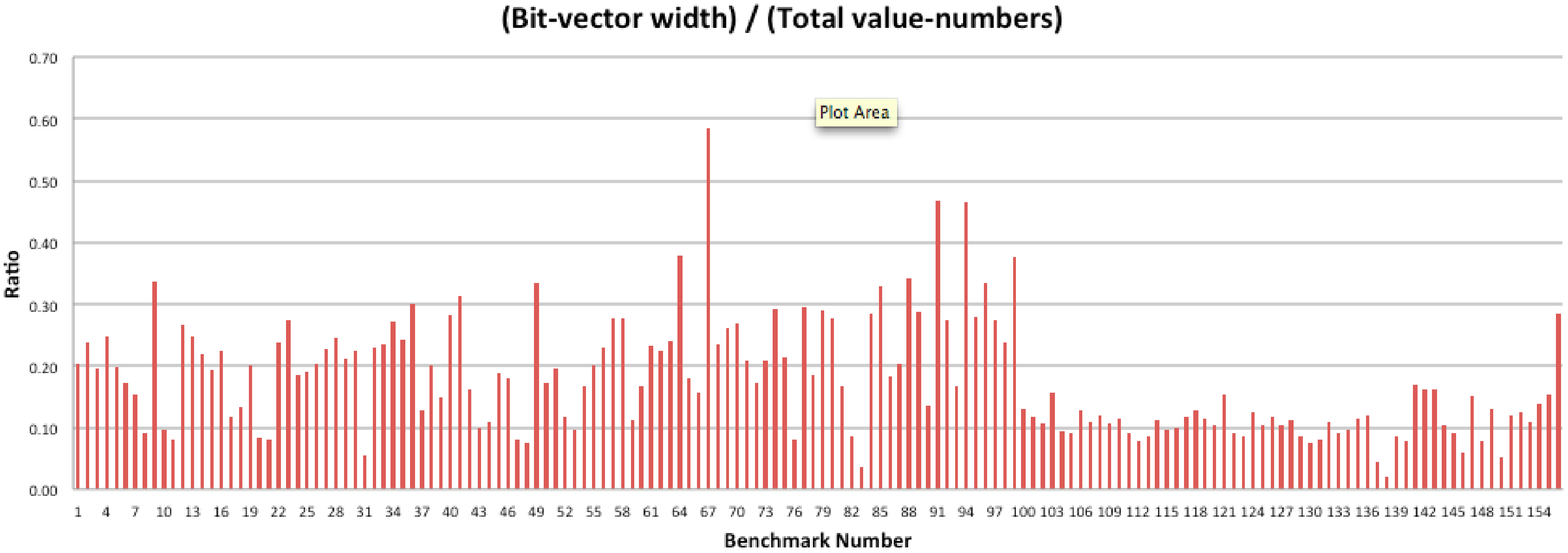} 
 \caption{\label{fig:barC}} 
\end{figure}

\begin{flushleft}
\textbf{\large{Value-Number driven code motion}}
\end{flushleft}
We initially implemented the flow equations from the Lazy Code Motion paper \cite{Knoop}.
This set included a total of 13 bit vectors for each basic block - 2 for block local
properties ANTLOC and TRANSP, and 11 for global properties. These equations,
           however, could only be applied to single instruction basic blocks.
           We therefore, derived a new set of equations which are motived by
           later work\cite{Knoop2} of the same authors.
           This set of equations apply to maximal basic blocks and
           entails a total of 19 bit vectors for each basic block in our
           current implementation - 3 for block local properties ANTLOC,
           TRANSP, XCOMP and 16 for global properties.  Block local properties are 
	  defined in appendix A. In appendix C, we include the generalized data flow 
	  framework, and show how each PRE equation maps to the framework. 
	  We call the algorithm value-number driven because each
           slot in each of the bit vectors is reserved for a particular value
           number rather than a particular expression. Also, we make the
           observation that a large number of expressions in the program only
           occur once, and are not useful for PRE. Therefore, to further optimize
           for space and time, we only give bit vector slots to value numbers
           which have more than one expression linked to them. A downside to this
approach is that we could miss opportunities for loop invariant code motion. As a
      solution, we extend the bit vector to include value numbers which have
      only a single expression linked to them but only if the expression is
      inside a loop. Note that we still exclude the cases where the expression
      is not part of a loop. Figure \ref{fig:barC} quantifies the savings we observe using functions from the LLVM
multi-source package. In the worst case scenario, the bit-vector width maintained by our 
algorithm has to be equal to the maximum value-number assigned by the value-numbering pass. However, as 
the results show, the average ratio of bit-vector width to maximum value-number is 0.18. This reflects a savings of
over 80\%.

\begin{flushleft}
\textbf{\large{Local CSE}}
\end{flushleft}
For our data flow equations to work efficiently, a local CSE pass is run
on each basic block. Basically, this pass removes the redundancies in
straight line basic block code and sanitizes it for the iterative bit vector
algorithm. Borrowed from \cite{Knoop2}, the main idea
is to trim the amount of work to be done by the PRE pass. For example, if there are many
expressions with the same value number in a basic block, rather than PRE going over all of them, 
local CSE can weed out the redundancies. We perform this step before calling
our data flow framework. 

\begin{flushleft}
\textbf{\large{Insert and Replace}}
\end{flushleft}
To maintain compatibility with SSA, we perform insertion and replacement
  through memory and re-run the \emph{mem2reg} pass after our PRE pass to convert the
  newly created load and store instructions to register operations. Following
  are the major points:
\begin{itemize}  
  \item Assign stack space \emph{(allocas)} at the beginning of the
  function for all the expressions that need movement.
  \item At insertion point,
           compute the expression and save the value to the stack slot assigned to the
             expression 
  \item At replacement point, load from the correct stack
             slot, replace all uses of the original expression with the load
             instruction, and delete the original expression
  \item \emph{mem2reg} converts stack operations to register operations and introduces the 
             necessary ${\Phi}$ instructions
\end{itemize}  

In appendix E, we have shown in Figure \ref{fig:1}
and Figure \ref{fig:2}, the optimizations performed by our PRE pass. 

\begin{flushleft}
\textbf{\Large{5. Miscellaneous}}
\end{flushleft}

\begin{flushleft}
\textbf{\large{Zero-trip Loops}}
\end{flushleft}
Our algorithm moves the loop invariant
      computations to the loop pre-header only if placement in the loop pre-header is 
      anticipatible. Such a pre-header is always available for \emph{do-while} loops, 
      but not for \emph{while} and \emph{for} loops. Hence, a modification is required
      to the structure of \emph{while} and \emph{for} loops which peels off the first
      iteration of the loop, protected by the loop condition. This alteration provides 
      PRE with a suitable loop pre-header to hoist loop independent computations to.
      In Figure \ref{fig:5}  (Appendix D) we show the CFG changes. We achieved this effect using an 
      existing LLVM pass \emph{-loop-rotate}.

\begin{flushleft}
\textbf{\large{Critical Edges}}
\end{flushleft}

A critical edge in a flow graph is an edge from a node with multiple successors to a node with multiple predecessors. Splitting such edges and inserting dummy nodes aids PRE by offering more anticipatible points. We used an existing LLVM pass (BreakCriticalEdges) for the same. In many cases, however, the dummy nodes created by this pass do not hold any computation after PRE. We used \emph{-simplifycfg} to clean up the mess created by BreakCriticalEdges.

\begin{flushleft}
\textbf{\large{Unresolved Issues}}
\end{flushleft}
There were a couple of issues on which we would have liked to spend more time.
The first is redundancy elimination for expressions which are lexically
different in SSA, and have different value numbers. We came up with a few
techniques within the bounds of our existing PRE code, but unfortunately, none
could be generalized to solve the core problem. The second issue pertains to
the insertion step of our algorithm and needs slightly detailed explanation.
Suppose that an expression, with value number vn, is to be inserted in a basic
block. Although our algorithm can handle all cases, for simplicity, assume that
the insertion point is the end of the basic block. To insert the expression we
scan the list of the expressions in the whole function which have the same
value number vn.  We then clone one of these expressions (called
    \emph{provider}) and place at the end of the basic block. The trivial case
is when the provider is available in the same basic block. If however, the
provider comes from another basic block, then we need to ensure that the
operands of the provider dominate the basic block we wish to insert the
expression in. Not being able to find a suitable provider is the only case
where we override the suggestion of the data flow analysis and not do PRE for
that expression only. PRE for other expressions proceeds as usual. Our
exhaustive testing on multiple suites suggests that this is a very rare
occurrence. 

\newpage  
\begin{flushleft}
\textbf{\Large{Testing}}
\end{flushleft}

While working on the project, we wrote 25 small test cases to capture the intricate movements
of expressions in the partial redundancy elimination algorithm. Most of these contrived test cases, along with our full source code, 
can be found on our project Github link \url{https://github.com/sdasgup3/PRE}. For evaluation on real life applications, we
chose 3 different suites - LLVM SingleSource, LLVM MultiSource, SPEC2006. For correctness, we checked the 
output of the binary optimized with our PRE pass with the provided reference output. All benchmarks passed the 
correctness test. For each suite, we present two sets of performance results. The first set compares the performance
of binaries optimized with our version of PRE (henceforth referred to as LCM-PRE) with binaries without PRE
optimization (henceforth referred to as BASE). The second set compares the
performance of LCM-PRE binaries with binaries optimized with LLVM's version of
PRE (henceforth referred to as GVN-PRE). To remove noise, we run each benchmark
thrice and take the average. Also, benchmarks with running time of less than 5
seconds are not accounted for. The next two subsections describe the
performance S-curves, following which we summarize in a table, the absolute
run-times for three benchmarks from each suite. For a meaningful comparison,
we use the same set of optimization knobs for BASE, GVN-PRE and LCM-PRE.

\begin{table}[h]
\centering
\begin{tabular}{|c|l|}
\hline
Pass Name & \qquad \quad \quad \quad \quad \quad \quad \quad \quad -opt switches                                                  \\ \hline \hline
\texttt{BASE}  & -mem2reg -loop-rotate -reassociate      -mem2reg -simplifycfg      \\ \hline
\texttt{LCM-PRE}   & -mem2reg -loop-rotate -reassociate -lcm -mem2reg -simplifycfg \\ \hline
\texttt{GVN-PRE}   & -mem2reg -loop-rotate -reassociate -gvn -mem2reg -simplifycfg \\ \hline
\end{tabular}
\caption{Optimization knobs.} \label{tab:1}
\end{table}

\begin{flushleft}
\textbf{\normalsize{LLVM Single source \& Multi source}}
\end{flushleft}

We ran $45$ benchmarks from the SingleSource package. Figures \ref{fig:6}(a)
shows the S-curve for BASE time over LCM-PRE time. For most of the
benchmarks (40/45) we either increase performance (up to 42\%) or maintain the
same level. $5$ benchmarks show slight degradation which is bound by 6.5\%.
Figure \ref{fig:6}(b) shows the S-curve for GVN-PRE time over LCM-PRE time. It is
heartening to beat GVN-PRE in a few cases. 

Results for the MultiSource benchmarks follow a similar trend. Out of the $45$ benchmarks
from this package, $41$ show improvement (up to 23\%) or maintain same
performance for BASE time over LCM-PRE time (Figure \ref{fig:7}(a)), while
degradation for the rest is bound by 5\%.
GVN-PRE time over LCM-PRE time is shown in \ref{fig:7}(b).

\begin{figure}
\centering
\begin{tabular}{c c}
  \scalebox{0.75}{
      \begin{tikzpicture}
      \pgfplotsset{every axis legend/.append style={
          at={(0.5,1.03)},
          anchor=south}}
      \begin{axis}[
        xlabel=Benchmark number,
        ylabel=Speedup,
        ymax=1.5, ymin=0.5, xmax=50,
        x tick label style={black},
        grid=both,xmajorgrids=false,
        ]
      \addplot table [y=BASE_LCM_SS, x=N]{data_ss.dat};
      \addlegendentry {$\text{Speedup } = \frac{BaseTime}{LcmTime}$}
      \end{axis}
      \end{tikzpicture}
  }
&
  \scalebox{0.75}{
      \begin{tikzpicture}
      \pgfplotsset{every axis legend/.append style={
          at={(0.5,1.03)},
          anchor=south}}
      \begin{axis}[
        xlabel=Benchmark number,
        ylabel=Speedup,
        ymax=1.5, ymin=0.5, xmax=50,
        x tick label style={black},
        grid=both,xmajorgrids=false,
        ]
    \addplot table [y=GVN_LCM_SS, x=N]{data_ss.dat};
    \addlegendentry {$\text{Speedup } = \frac{GvnTime}{LcmTime}$}
      \end{axis}
      \end{tikzpicture}
}
\\
\qquad (a) & \quad (b) \\    
\end{tabular}
\caption{Performance evaluation with LLVM SingleSource Benchmark}
\label{fig:6}
\end{figure}
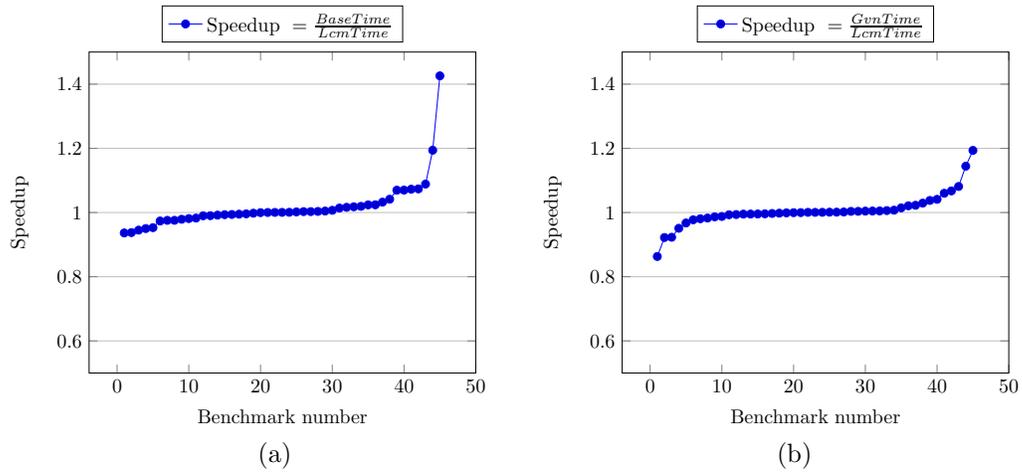

\begin{figure}
\begin{tabular}{c c}
  \scalebox{0.75}{
      \begin{tikzpicture}
      \pgfplotsset{every axis legend/.append style={
          at={(0.5,1.03)},
          anchor=south}}
      \begin{axis}[
        xlabel=Benchmark number,
        ylabel=Speedup,
        ymax=1.5, ymin=0.5, xmax=50,
        x tick label style={black},
        grid=both,xmajorgrids=false,
        ]
      \addplot table [y=BASE_LCM_MS, x=N]{data_ms.dat};
      \addlegendentry {$\text{Speedup } = \frac{BaseTime}{LcmTime}$}
      \end{axis}
      \end{tikzpicture}
  }
&
  \scalebox{0.75}{
    \begin{tikzpicture}
      \pgfplotsset{every axis legend/.append style={
          at={(0.5,1.03)},
          anchor=south}}
    \begin{axis}[
      xlabel=Benchmark number,
      ylabel=Speedup,
      ymax=1.5, ymin=0.5, xmax=50,
      x tick label style={black},
      grid=both,xmajorgrids=false,
    ]
    \addplot table [y=GVN_LCM_MS, x=N]{data_ms.dat};
    \addlegendentry {$\text{Speedup } = \frac{GvnTime}{LcmTime}$}
\end{axis}
\end{tikzpicture}
  } \\
\qquad (a) & \quad (b) \\    
\end{tabular}
\caption{Performance evaluation with LLVM MultiSource Benchmark}
\label{fig:7}
\end{figure}
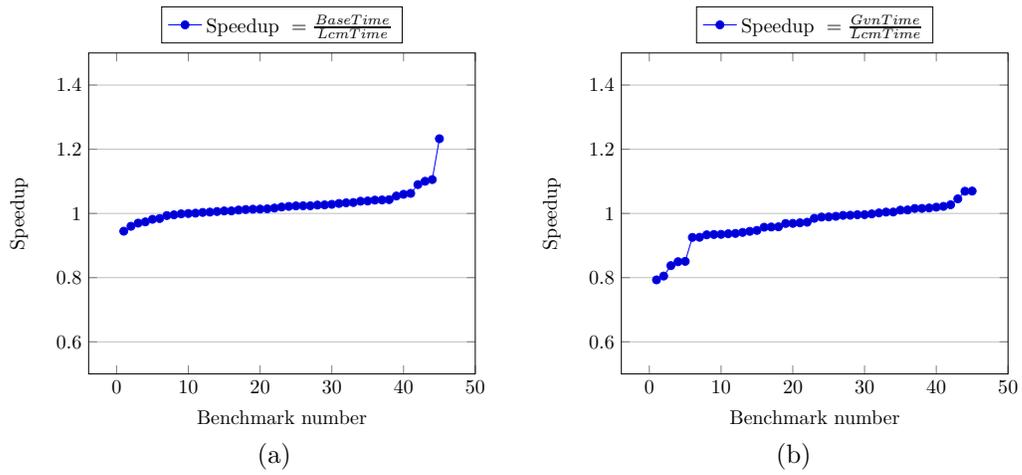

\begin{flushleft}
\textbf{\normalsize{Spec2006 Benchmark}}
\end{flushleft}
We augmented our testing infrastructure to support the SPEC2006 suite. Both
SPEC-INT and SPEC-FP were tested. We, however, had to limit our testing to C/C++
benchmarks, and leave out Fortran. Getting SPEC-Fortran benchmarks to run
inside LLVM needs extra support. We take our inputs for the SPEC runs from the
following source -
\url{http://boegel.kejo.be/ELIS/spec_cpu2006/spec_cpu2006_command_lines.html}

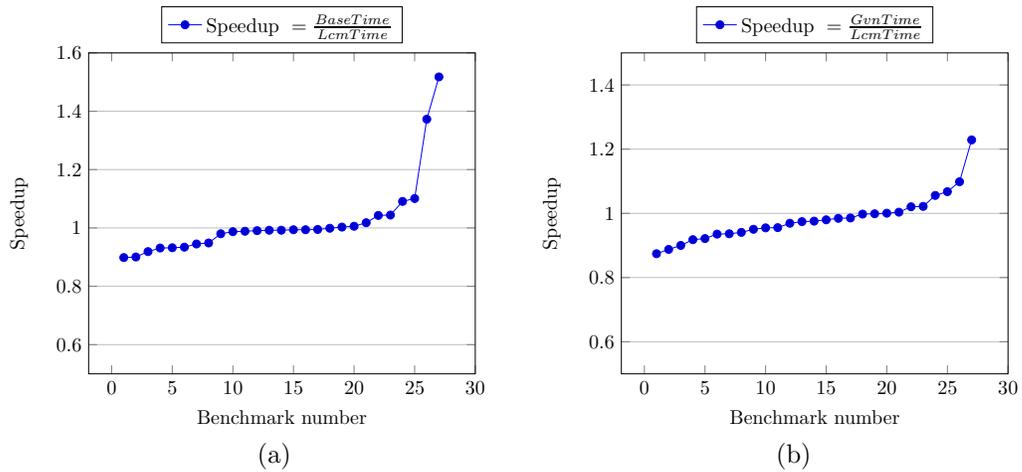
\begin{figure}
\begin{tabular}{c c}
  \scalebox{0.75}{
      \begin{tikzpicture}
      \pgfplotsset{every axis legend/.append style={
          at={(0.5,1.03)},
          anchor=south}}
      \begin{axis}[
        xlabel=Benchmark number,
        ylabel=Speedup,
        ymax=1.6, ymin=0.5, xmax=30,
        x tick label style={black},
        grid=both,xmajorgrids=false,
        ]
      \addplot table [y=BASE_LCM, x=N]{data_spec_int.dat};
      \addlegendentry {$\text{Speedup } = \frac{BaseTime}{LcmTime}$}
      \end{axis}
      \end{tikzpicture}
  }
&
  \scalebox{0.75}{
    \begin{tikzpicture}
      \pgfplotsset{every axis legend/.append style={
          at={(0.5,1.03)},
          anchor=south}}
    \begin{axis}[
      xlabel=Benchmark number,
      ylabel=Speedup,
      ymax=1.5, ymin=0.5, xmax=30,
      x tick label style={black},
      grid=both,xmajorgrids=false,
    ]
    \addplot table [y=GVN_LCM, x=N]{data_spec_int.dat};
    \addlegendentry {$\text{Speedup } = \frac{GvnTime}{LcmTime}$}
\end{axis}
\end{tikzpicture}
  }
\\
\qquad (a) & \quad (b) \\    
\end{tabular}
\caption{Performance evaluation with SPEC 2006}
\label{fig:8}
\end{figure}

Out of the $27$ runs from SPEC, $19$ show improvement (up to 52\%) or
maintain same performance for BASE time over LCM-PRE time, while
degradation for the rest is bound by 10\% (Figure \ref{fig:8}(a)). 
Our pass triumphs over GVN-PRE for quite a few cases here
as well as shown in \ref{fig:8}(b). 

\begin{table}
\centering
\scalebox{.85}{
  \begin{tabular}{|l|l|l|l|l|l|}
  \hline 
  \multicolumn{1}{|c|}{Benchmark Name}          & \begin{tabular}[c]{@{}l@{}}BASE \\ Time (B)\end{tabular} & \begin{tabular}[c]{@{}l@{}}LCM-PRE \\ Time(L)\end{tabular} & \multicolumn{1}{c|}{\begin{tabular}[c]{@{}c@{}}GVN-PRE \\ Time(G)\end{tabular}} & B/L   & G/L   \\ \hline \hline
  \texttt{SingleSource/Benchmarks/Dhrystone/fldry}       & 5.405                                                        & 4.965                                                      & 5.263                                                                           & 1.088 & 1.060 \\ \hline
  \texttt{SingleSource/Benchmarks/Misc/oourafft}         & 5.112                                                        & 4.281                                                      & 4.286                                                                           & 1.194 & 1.001 \\ \hline
  \texttt{SingleSource/Benchmarks/Misc/lowercase}        & 40.795                                                       & 28.612                                                     & 28.628                                                                          & 1.425 & 1.000 \\ \hline
  \texttt{MultiSource/Benchmarks/TSVC/NodeSplitting-flt} & 7.378                                                        & 6.706                                                      & 5.398                                                                           & 1.100 & 0.804 \\ \hline
  \texttt{MultiSource/Benchmarks/TSVC/Expansion-flt}     & 6.339                                                        & 5.734                                                      & 5.308                                                                           & 1.105 & 0.925 \\ \hline
  \texttt{MultiSource/Benchmarks/TSVC/Expansion-dbl}     & 7.134                                                        & 5.787                                                      & 5.355                                                                           & 1.232 & 0.925 \\ \hline
  \texttt{SPECINT2006/456.hmmer (ref-input 1)}                         & 1547.77     & 1019.885  & 993.581      & 1.517 & 0.974 \\ \hline
  \texttt{SPECINT2006/456.hmmer (ref-input 2)}                         &  715.641    &  521.318  & 455.754      & 1.372   & 0.874   \\ \hline
  \texttt{SPECINT2006/464.h264ref}                       & 185.551     & 168.563   & 163.363      & 1.100 & 0.969 \\ \hline
  \end{tabular}
}
\caption{Absolute run-times for $3$ benchmarks from each suite} \label{tab:2}
    \end{table}

\newpage
\begin{flushleft}
\textbf{\large{{Performance Analysis }}}
\end{flushleft} 
We analyzed the benchmarks where our pass degrades performance. In this
subsection, we summarize our thoughts and findings. To measure the improvements
for LCM-PRE over BASE, we switch off all backend optimizations for all runs.
More specifically, we use -O$0$ while converting the LLVM bitcode to machine
code. A major repercussion of using -O$0$ is that none of the efficient register
allocators (greedy, pbqp) can be used (LLVM restriction). Hence, we were
stuck with the fast register allocator which does a very poor job for some of
the benchmarks. The performance of LCM-PRE is sensitive to register allocation
(because of increased register pressure), and this causes the performance dip
over BASE as presented in the S-curves. We substantiate this claim with an
example from the SingleSource package \emph{(Benchmarks/Shootout-C++/methcall)} (Table
    \ref{tab:3}). Data from llc-dump shows the increased amount of loads and
stores to the stack for LCM-PRE. We also gather the dynamic data from Pin,
       using a simple opcode-mix tool which we wrote. The increased number of
       stack reads (15\%) and writes (6\%) at runtime for LCM-PRE confirms our
       hypothesis. We hold the opinion that using a more powerful register
       allocator would wipe off most of the performance losses. 

Next we explain why we chose to stick with -O$0$ rather than using -O$3$. This was
done to disable backend optimizations such as \texttt{-machine-licm}. We expect major
performance gains from the loop invariant code motion done by LCM-PRE, and
allowing a backend pass to achieve the same effect on BASE would steal our
thunder. This was confirmed experimentally, where using -O$3$ in the backend
results in LCM-PRE execution time same as BASE for all the benchmarks (no
    improvement, no degradation).

\begin{table}[h]
\centering
\begin{tabular}{|c|c|c|}
\hline
\multicolumn{3}{|c|}{Dynamic data from Pin tool}                                                          \\
\hline
\hline
\begin{tabular}[c]{@{}c@{}}Stats \end{tabular} 			& LCM-PRE & BASE                \\ \hline
Total Instructions (in Billion)                                                     & 72      & 68                      \\ \hline
stack-read count (in Billion)                                                 & 19      & 17                      \\ \hline
stack-write count (in Billion)                                                 & 16      & 15                      \\ \hline \hline
\multicolumn{3}{|c|}{Static data from llc tool - fast register allocator}                                                               \\ \hline
\multicolumn{1}{|l|}{regalloc-Number of loads added}        & 34      & \multicolumn{1}{c|}{28} \\ \hline
\multicolumn{1}{|l|}{regalloc-Number of stores added}       & 35      & \multicolumn{1}{c|}{30} \\ \hline
\end{tabular}
\caption{\label{tab:3}}
\end{table}

\cmnt{
Apart from measured runtime on the hardware, another metric to quantify the
effects of our optimization pass is the dynamic instruction count. We use Pin
(dynamic binary instrumentation tool) from Intel for this purpose. We have
written a very simple Pintool to dump the dynamic instruction count of each
type of instruction. We would present supporting data from Pin our final
report.

\begin{flushleft}
\textbf{\Large{{Conclustion and Future Work }}}
\end{flushleft} 
In this phase, we were able to code the missing pieces of our PRE algorithm,
   thereby achieving full functionality of the project. The highlights were the
   insert-replace algorithm, LICM improvements, and fixes for numerous bugs
   which surfaced during testing on the LLVM single-source package. We have
   written scripts to automate the testing, and this would speed up work for
   the final phase. \\ 
   The S-curve in this report (Figure 5)
   presents improvements with respect to the baseline (no PRE pass) for
   single-source package. We would like to evaluate the cases where our pass
   degrades performance compared to baseline. In our final report, we plan to
   include similar curves for benchamarks from the LLVM multi-source package as
   well as the SPEC 2006 suite. Also, we would include S-curves to compare the
   effectiveness of our PRE pass with the LLVM GVN-PRE pass. For extreme
   outliers, we hope to present supporting data to reason about the performance
   change. Pin tool analysis and the statistics dumped by our PRE pass would be
   used for this.
}

\input{appendix_report.tex}

\nocite{*}
\bibliography{report}
  
\end{document}

%% file: appendix_report.tex
\appendix

\chapter{Computation of localized sets}
For each basic block there are 3 bit vectors dedicated to the block-specific
properties, namely \texttt{Transp}, \texttt{Antloc} and \texttt{Xcomp}. As
mentioned before, a bit vector is a boolean array of value numbers.  Let the
leader expression (as defined in the section on value numbering) associated
with the value number $v$ be called $L(v)$.

\begin{equation}
\begin{array}{l c l}
\texttt{Transp(v,B)} &=& \left\{
                    \begin{array}{l l}
                        false & \quad \text{iff $\exists$ x $\in$ operands of L(v) such that \texttt{Mod(x,B)} = true}\\
                        true & \quad \text{Otherwise}
                    \end{array} \right. \\
\texttt{Antloc(v,B)} &=& \texttt{Eval(v,B)} \cap \texttt{Transp(v,B)} \\
\texttt{Xcomp(v,B)} &=& \texttt{Eval(v,B)} \cap \overline{\texttt{Transp(v,B)}}\\
  &&\\
\text{where}&& \\  
\quad \texttt{Eval(v,B)} &=& \text{\{v $|$ value number v is computed in B\}}\\
\quad \texttt{Mod(op,B)} &=& \text{operand op modified in B}\\
\end{array}
\end{equation}

\chapter{Lazy Code motion Transformations}

\begin{itemize}
\item Down Safety Analysis (Backward data flow analysis)
\begin{equation}
\begin{array}{l c l}
\antin{b} &=& \antloc{b} \cup (\transp{b} \cap \antout{b}) \\
\antout{b} &=& \xcomp{b} \cup \left\{
                    \begin{array}{l l}
                        \phi & \quad \text{if b = exit}\\
                        \displaystyle \bigcap_{s \in succ(b)} \antin{s} &
                    \end{array} \right. \\
\end{array}
\end{equation}

\item Up Safety Analysis (Forward data flow analysis)
\begin{equation}
\begin{array}{l c l}
\availin{b} &=& \left\{
                  \begin{array}{l l}
                        \phi & \quad \text{if b = entry}\\
                        \displaystyle \bigcap_{p \in pred(b)} (\xcomp{p} \cup \availout{p}) & 
                  \end{array} 
              \right. \\
\availout{b} &=& \transp{b} \cap (\antloc{b} \cup \availin{b}) \\
\end{array}
\end{equation}

\item Earliest-ness (No data flow analysis)
\begin{equation}
\begin{array}{l c l}
\earlin{b}  &=& \antin{b} \cap \displaystyle \bigcap_{p \in pred(b)} (\overline{\availout{p} \cup \antout{p}}) \\ 
\earlout{b} &=& \antout{b} \cap \overline{\transp{b}}
\end{array}
\end{equation}

\item Delayability (Forward data flow analysis)
\begin{equation}
\begin{array}{l c l}
\delayin{b} &=& \earlin{b} \cup  \left\{
                              \begin{array}{l l}
                                \phi & \quad \text{if b = entry}\\
                                \displaystyle \bigcap_{p \in pred(b)} (\overline{\xcomp{p}} \cap \delayout{p}) & 
                  \end{array} 
              \right. \\
\delayout{b} &=& \earlout{b} \cup (\delayin{b} \cap \overline{\antloc{b}}) \\
\end{array}
\end{equation}

\item Latest-ness (No data flow analysis)
\begin{equation}
\begin{array}{l c l}
\latestin{b}  &=& \delayin{b} \cap \antloc{b}\\
\latestout{b} &=& \delayout{b} \cap (\xcomp{b} \cup \displaystyle \bigcup_{s \in succ(b)} \overline{\delayin{s}})
\end{array}
\end{equation}

\item Isolation Analysis (Backward data flow analysis)
\begin{equation}
\begin{array}{l c l}
\isoin{b} &=& \earlout{b} \cup \isoout{b} \\
\isoout{b} &=& \left\{
                    \begin{array}{l l}
                        U & \quad \text{if b = exit}\\
                        \displaystyle \bigcap_{s \in succ(b)} (\earlin{s} \cup (\overline{\antloc{s}} \cap \isoin{s}) )&
                    \end{array} \right. \\
\end{array}
\end{equation}

\item Insert and Replace points
\begin{equation}
\begin{array}{l c l}
\insertin{b} &=& \latestin{b} \cap \overline{\isoin{b}} \\
\insertout{b} &=& \latestout{b} \cap \overline{\isoout{b}} \\
&&\\
\replacein{b} &=& \antloc{b} \cap \overline{\latestin{b} \cap \isoin{b}} \\
\replaceout{b} &=& \xcomp{b} \cap \overline{\latestout{b} \cap \isoout{b}}
\end{array}
\end{equation}
\end{itemize}

\chapter{Generalized data flow framework}

All the equations in Appendix B can be computed using the generic
framework defined below.

\section{Forward Analysis}
\begin{equation}
\begin{array}{l c l}
\myin{b} &=& \Alpha{b} \cup  \left\{
                    \begin{array}{l l}
                        \bot & \quad \text{if b = entry}\\
                        \displaystyle \bigwedge_{p \in pred(b)} \Beta{p}&
                    \end{array} \right. \\
\myout{b} &=& \myGamma{b}                      
\end{array}
\end{equation}

\section{Backward Analysis}
\begin{equation}
\begin{array}{l c l}
\myin{b} &=& \myGamma{b}                      \\
\myout{b} &=& \Alpha{b} \cup  \left\{
                    \begin{array}{l l}
                        \bot & \quad \text{if b = exit}\\
                        \displaystyle \bigwedge_{s \in succ(b)} \Beta{s}&
                    \end{array} \right. \\
\end{array}
\end{equation}

The following is the function which we call with dataflow equation
specific parameters defined subsequently.

\begin{framed}
\[
\texttt{callFramework}(\myout{b}, \myin{b}, \Alpha{b}, \Beta{b},\myGamma{b}, \bigwedge, \bot, \top, \text{Direction})
\]
\end{framed}

Following is the list of values that we need to plug-in to $\alpha$,
          $\beta$ and $\gamma$ for the above generic framework
          to work.

\begin{itemize}
\item Down Safety Analysis (Backward data flow analysis)
\begin{equation}
\begin{array}{l c l}
\Alpha{x}     &=& \xcomp{x} \\
\Beta{x}      &=& \antin{x}     \\     
\myGamma{x}   &=& \transp{x} \cap \antout{x} \cup \antloc{x}\\
\bigwedge     &=&  \cap \\
\bot          &=& \phi \\
\top          &=& V, \text{set of all values} \\
\text{Direction}    &=& \text{Backward}
\end{array}
\end{equation}

\item Up Safety Analysis (Forward data flow analysis)
\begin{equation}
\begin{array}{l c l}
\Beta{x}      &=& \xcomp{x} \cup \availout{x}     \\     
\myGamma{x}   &=& \antloc{x} \cup \availin{x} \cap \transp{x}\\
\bigwedge     &=&  \cap \\
\bot          &=& \phi \\
\top          &=& V, \text{set of all values} \\
\text{Direction}    &=& \text{Forward}
\end{array}
\end{equation}

\item Delayability (Forward data flow analysis)
\begin{equation}
\begin{array}{l c l}
\Alpha{x}     &=& \earlin{x} \\
\Beta{x}      &=& \overline{\xcomp{x}} \cap \delayout{x}     \\     
\myGamma{x}   &=& \delayin{x} \cap \overline{\antloc{x}} \cup \earlout{x}\\
\bigwedge     &=&  \cap \\
\bot          &=& \phi \\
\top          &=& V, \text{set of all values} \\
\text{Direction}    &=& \text{Forward}
\end{array}
\end{equation}

\item Isolation Analysis (Backward data flow analysis)
\begin{equation}
\begin{array}{l c l}
\Beta{x}      &=& \overline{\antloc{x}} \cap \isoin{x} \cup \earlin{x}     \\     
\myGamma{x}   &=& \earlout{x} \cup \isoout{x} \\
\bigwedge     &=&  \cap \\
\bot          &=& V, \text{set of all values} \\
\top          &=& V, \text{set of all values} \\
\text{Direction}    &=& \text{Backward}
\end{array}
\end{equation}

\end{itemize}

\chapter{Transformations for ``Zero-trip Loops''}

\begin{figure}[htbp]
  \begin{center}
     \includegraphics[scale=0.5]{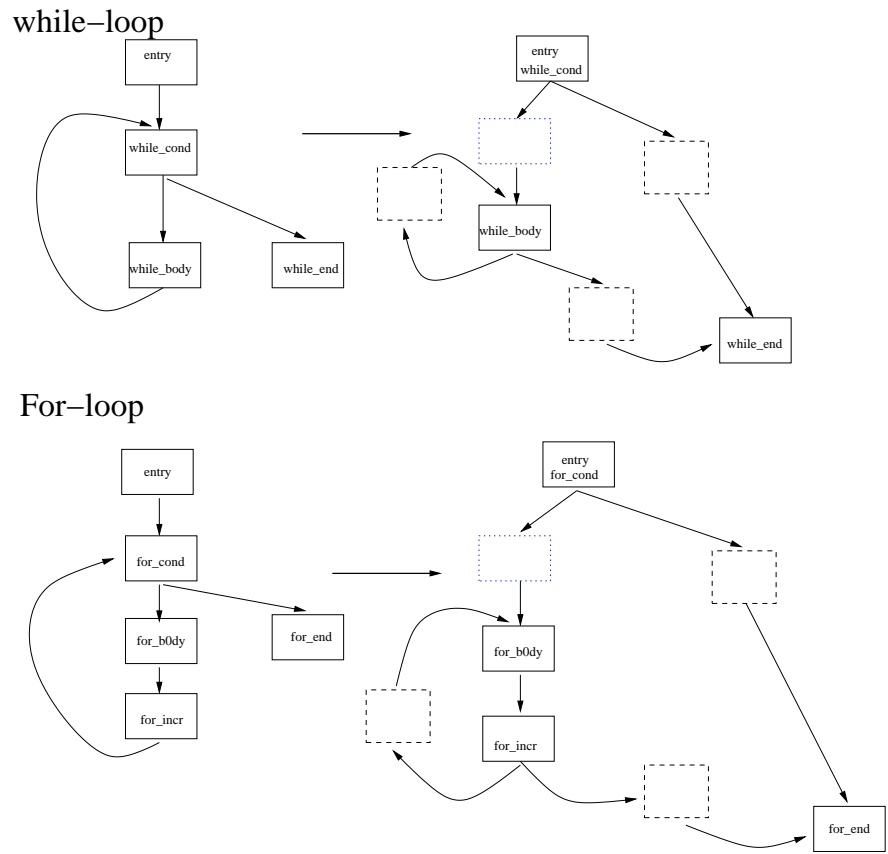} 
  \end{center}
  \caption{Loop transformations done by \emph{-loop-rotate}. \emph{do-while}
    loops remain unaffected. Blue dotted boxes are the ones inserted by loop
      rotate. PRE can insert the computations in these places.}
  \label{fig:5} 
  \end{figure}

\chapter{An Extended Example}

\begin{figure}[htbp]
  \begin{center}
     \includegraphics[scale=0.3]{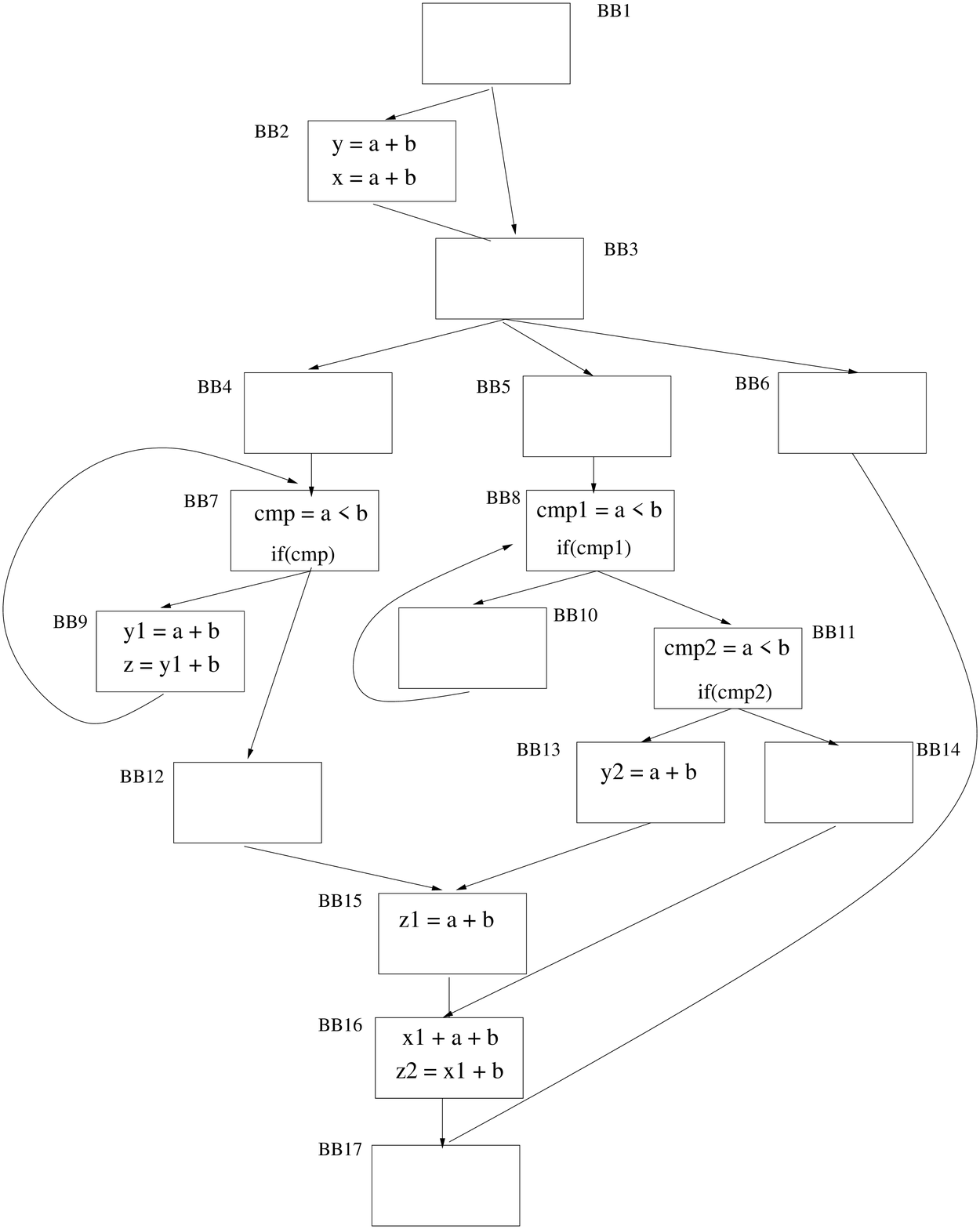} 
  \end{center}
  \caption{A motivating example}
    \label{fig:1} 
\end{figure}

\begin{figure}[htbp]
  \begin{center}
     \includegraphics[scale=0.3]{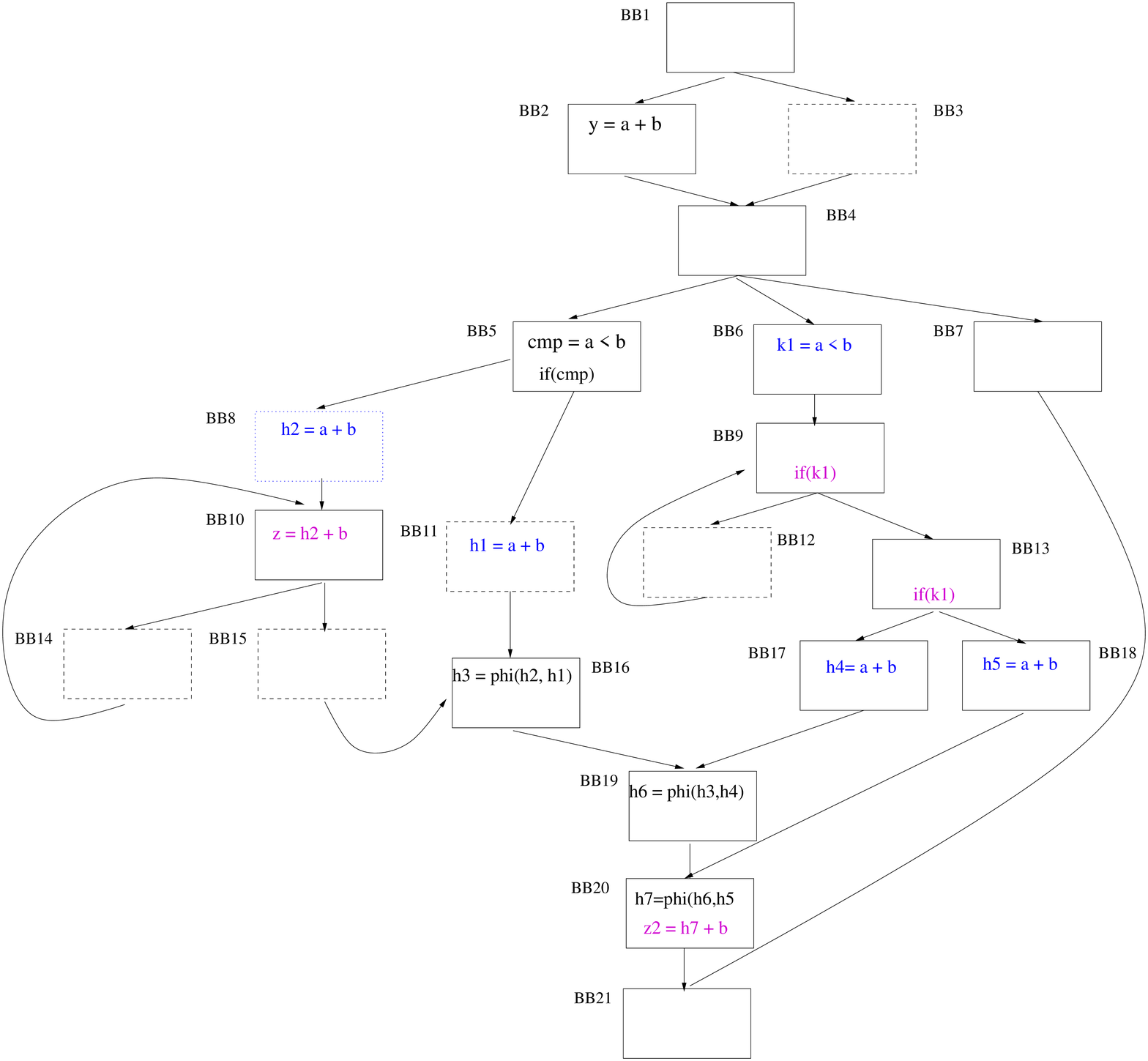} 
  \end{center}
  \caption{Lazy code motion transformation on computations $a + b$ \& $a < b$.}
  \label{fig:2} 
\end{figure}

Here we show, through an example code, the optimizations performed
by our PRE pass. The intention here is to highlight redundancy elimination for  
expressions $a + b$ \& $a < b$. Optimal placements are marked in Figure \ref{fig:2}.
Some of the notable obseravtions are:
\begin{itemize}
\item Black dotted boxes denote basic blocks inserted because of critical edge splitting
\item  Blue dotted boxes are the loop pre-headers inserted
      by -loop-rotate pass. PRE can insert computations here.
\item Inserted statements are marked \textcolor{blue}{blue} and replaced ones with
      \textcolor{magenta}{magenta}      
\item LCSE (Local common subexpression elimination)  happened in BB2.
\item For the loop {BB7,BB9} in Figure \ref{fig:1}, LICM happened wherein the computation 
of $a+b$ is moved from BB9 (in Figure  \ref{fig:1}) to BB8 (in Figure \ref{fig:2}). BB8 is the loop pre-header
\end{itemize}

